\documentclass[showpacs,aps,pre,twocolumn,superscriptaddress,textsuperscript]{revtex4}
\usepackage{graphicx}
\usepackage{amssymb}
\usepackage[sort&compress]{natbib}
\begin{document}
\title {Atomically thin mononitrides SiN and GeN: new two-dimensional semiconducting materials}
\author{Yan Qian}
\email[Corresponding author:]{qianyan@njust.edu.cn}
\affiliation{Department of Applied Physics, Nanjing University of Science and Technology, Nanjing 210094, China}
\author{Zhengwei Du}
\affiliation{Department of Applied Physics, Nanjing University of Science and Technology, Nanjing 210094, China}
\author{Renzhu Zhu}
\affiliation{Department of Applied Physics, Nanjing University of Science and Technology, Nanjing 210094, China}
\author{Haiping Wu}
\email[Corresponding author:] {mrhpwu@njust.edu.cn}
\affiliation{Department of Applied Physics, Nanjing University of Science and Technology, Nanjing 210094, China}
\author{Erjun Kan}
\affiliation{Department of Applied Physics, Nanjing University of Science and Technology, Nanjing 210094, China}
\author{Kaiming Deng}
\affiliation{Department of Applied Physics, Nanjing University of Science and Technology, Nanjing 210094, China}

\date{\today}
\begin{abstract}
Low-dimensional Si-based semiconductors are unique materials that can both match well with the Si-based electronics and satisfy the demand of miniaturization in modern industry. Owing to the lack of such materials, many researchers put their efforts into this field. In this work, employing a swarm structure search method and density functional theory, we theoretically predict two-dimensional atomically thin mononitrides SiN and GeN, both of which present semiconducting nature. Furthermore study shows that SiN and GeN behave as indirect band gap semiconductors with the gap of 1.75 and 1.20 eV, respectively. The ab initio molecular dynamics calculation tells that both two mononitrides can exist stably even at extremely high temperature of 2000 K. Notably, electron mobilities are evaluated as $\sim$0.888$\times$10$^{3}$cm$^{2}$V$^{-1}$s$^{-1}$ and $\sim$0.413$\times$10$^{3}$cm$^{2}$V$^{-1}$s$^{-1}$ for SiN and GeN, respectively. The present work expands the family of low-dimensional Si-based semiconductors.
\end{abstract}
\maketitle
\section{Introduction}
With the development of modern industry, devices with low power consumption and small scale are desirable. This demand attracts a lot of research focused on searching for corresponding materials. Two-dimensional (2D) materials are the potential ones that could be utilized in the future components\cite{Boon,Tang,Rao}. Fortunately, motivated by the exfoliation of graphene\cite{Novoselov1,Novoselov2}, researchers put a great of their work into designing new 2D materials or tuning properties of the existing 2D materials via theoretical or experimental methods. For instance, h-BN sheet, borophene, phosphorene, germanene, silicene, 2D transition metal dichalcogenides and 2D transition metal dinitride are reported in succession\cite{Pacil, Andrew, Cahangirov, Ding, Liu, Xu1, Chen, Fang, Wu}. A lot of methods, such as strain, vacancy, electric field, and so on are employed to regulate the properties of 2D materials\cite{Peng, Qian, Kan, HP}. On the other hand, some properties of low-dimensional materials are usually very different from the bulk counterparts. For instance, graphite behaves as common metallic nature, while graphene shows a zero-gap semiconductor with Dirac points\cite{Novoselov1,Novoselov2}. Their amazing properties also garner researchers' tremendous interest to design new functionalized components.

It is well known that semiconductors are the key materials for the modern industry, such as in integrated circuit, power component, light-emitting diode (LED) fields and so on. Associated with the demand of miniaturization of components, searching for 2D semiconducting materials shows so urgent. Benefitting from the extensive study on 2D materials, some literatures have reported on 2D semiconducting materials. For example, Hexagonal boron nitride (h-BN) behaves as a semiconductor with a gap of 5.8 eV\cite{Rubio}, monolayered MoS$_{2}$ shows semiconducting nature with a direct-gap of $\sim$1.80 eV\cite{Kin}, few-layer black phosphorus was also reported as a direct-gap semiconductor and the gap could increase up to 2.0 eV with decreasing the number of layers down to one\cite{Tran}. Monochalcogenides are a new class of 2D semiconductors (eg. GaS, GaSe, and GaTe) and have also attracted a great of attention\cite{Late1,Hu1,Zhou,Hu2,Xu2}. The electronic studies showed that 2D GaS and GaSe are indirect band gap semiconductors, while 2D GaTe is a direct band gap semiconductor. Furthermore, these 2D materials have been utilized to design field-effect transistors(FETs), highly sensitive phototransistors, and so on\cite{Late2, Hu2}. Another highlighted monochalcogenide is InSe\cite{Kress}. Notably, its 2D counterpart multilayer InSe was experimentally exfoliated or deposited on Si/SiO2 substrates\cite{Mudd,Gisbert}, and the further research revealed that the multilayer InSe has high electron mobility of $\sim$1055 cm$^{2}$V$^{-1}$s$^{-1}$ in InSe-based FETs\cite{Feng}, this value is much larger than that of TMD-based FETs\cite{YZhang}. In 2016, 2D few-layer InSe was encapsulated in h-BN by Bandurin et. al, and the density functional theory (DFT) study demonstrated that few-layer InSe behaves as an indirect bang gap semiconductor\cite{Bandurin}.

However, despite many 2D semiconductors have been reported by theoretical or experimental methods, few literatures were reported on applications of these 2D semiconductors, especially in integrated circuit. We think the reason is that, in the previous integrated circuit, semiconducting materials are usually grown on Si or SiC substrates, while so far reported 2D semiconductors can not match well with these substrates.

The above fact motivates researchers to search for 2D silicon-based semiconductors. Silicene, as a 2D silicon material with a buckling hexagonal honeycomb structure, was theoretically designed in 2009\cite{Cahangirov}. The electronic calculation told it exhibits gapless semimetal, instead of semiconductor like its bulk counterpart, and its charge carriers behave as a massless Dirac fermion due to linear band crossing at the Fermi energy level (\textit{E}$_{F}$). Subsequently, some work was focused on tuning the band gap of silicene. For example, Du \textit{et al} regulated silicene from semimetal to semiconductor by chemically adsorbing oxygen atoms, and the band gap can range from 0.11 to 0.30 eV through the adsorption configurations and amount of adsorbed oxygen atoms\cite{Yi}. But whether silicene with the oxygen atoms adsorbed on surface could match well with other silicon-based devices is an open question, because the adsorbed atoms would destroy the structure at the atomic level. Additionally, chemical modification is still hard to achieve by experiment now. Besides, Ni \textit{et al} theoretically predicted that silicene also could be tuned to a semiconductor via a vertical electric field\cite{Ni}. However, the electric field applied to open an effective gap is too high to achieve in experiment, especially in practical applications, for example, an electric field of 0.16 V/{\AA} is required to open a gap of 0.026 eV.

Therefor, searching for 2D silicides with intrinsic semiconducting nature, instead of tuning silicene by different means, is desired and maybe is a suitable way. Usually, compounds without unpaired electrons exhibits semiconducting character. It is well known that silicene is a semimetal with a buckling graphene-like structure, but it is instable in air, caused by the existence of dangling bond due to the \textit{sp}$^{3}$ hybridization of Si atom. Based on this fact, making monolayered SiN by substituting N for Si atoms in one Si sublattice of silicene and then combining two SiN monolayers via forming Si-Si bond could just eliminate the Si dangling bonds. Following this idea, we theoretically design a new 2D layered silicide SiN. Since Ge is the same group of Si, GeN with the similar structure is studied as well. As expected, the results demonstrate that 2D layered SiN and GeN both behave as semiconductors with band gaps of 1.75 and 1.20 eV, respectively.

\section{Computational methods}
In order to confirm our assumption, CALYPSO code is employed to search for the ground-state of 2D SiN and GeN firstly. The code is very one that developed to search for the stable structures of compounds by using the swarm-intelligence based structural prediction calculations\cite{wang1,wang2}. The underlying ab initio structural relaxations and electronic band structure calculations are carried out in the framework of DFT within generalized$-$gradient approximations using the Perdew$-$Burke$-$Ernzerhof (PBE) exchange$-$correlation functional and projector augmented wave (PAW) potentials\cite{Perdew, Kresse1} as implemented in VASP code\cite{Kresse2}. The structural relaxations are performed until the Hellmann-Feynman force on each atom is less than 0.001 eV/{\AA}. To ensure high accuracy, the k-point density and the plane waves cutoff energy are increased until the change of the total energy is less than 10$^{-5}$ eV, and the Brillouin$-$zone (BZ) integration is carried out using 15$\times$15$\times$1 Monkhorst$-$Pack grid in the first BZ, the plane waves with the kinetic energy up to 600 eV is employed. In addition, the simulations are performed using a 2$\times$2$\times$1 supercell based on unit cell, and the repeated layered geometry is with a thick vacuum region of 20 {\AA}. The phonon calculations are performed using a supercell approach implemented in the PHONOPY code\cite{K,A}.

\section{Results and discussions}
The structures of ground-state SiN and GeN are pictured in Fig.1, and some structural parameters are listed as well. The figure displays that the structures are just the ones as we expected. For the structure of layered SiN as pictured in Figs. 1(a) and 1(c), the bond length of Si-N is $\sim$1.76 {\AA}, this value is consistent with $\sim$1.72 or $\sim$1.73 {\AA} in the bulk Si$_{3}$N$_{4}$ given by experiment\cite{Hardie} and our calculation. The Si-Si bond length (i.e. the distance between two SiN atomic planes) is $\sim$2.43 {\AA}, which is somewhat larger than $\sim$2.34 and $\sim$2.25 {\AA} in single crystal Si and Silicene\cite{Cahangirov}. This slight change of bond length is originated from the different electronegativity between N and Si atoms. The electronegativity of N is stronger than that of Si atom, leading to the fact that there would be more electrons to form Si-N bond and less electrons to form Si-Si bond compared with the corresponding bonds in two pure Si isomers. Lessening of electrons would weaken Coulomb attractive force between Si atoms, resulting in the lengthening of bond. It is opposite to the case with electrons increased. Due to the \textit{sp}$^{3}$ hybridization of Si atom, the buckling characteristic present in SiN, and the buckling $\Delta$Z is $\sim$0.56 {\AA}, a bit larger than $\sim$0.45 {\AA} in pure silicene\cite{Cahangirov}. This could be explained by the following mechanism. With substituting N for Si atom in one Si sublattice, Si-N bond length is much shorter than that of Si-Si bond in silicene, i.e. the 2D atomic plane is heavily compressed, this plane compression pushes the N and Si sublayers outwards the plane and draws the enhancement of buckling. For the layered GeN plotted in Figs. 1(b) and 1(d), it has the similar structural configuration of layered SiN. In detail, the Ge-N and Ge-Ge bond lengthes are $\sim$1.91 and $\sim$2.57 {\AA}, respectively, and the buckling $\Delta$Z is $\sim$0.67 {\AA}. These lengthes of bonds are all larger than those of the corresponding bonds in 2D layered SiN, arisen from the larger radius of Ge than Si atoms.

\begin{figure}[htbp]
\centering
\includegraphics[width=8.5cm]{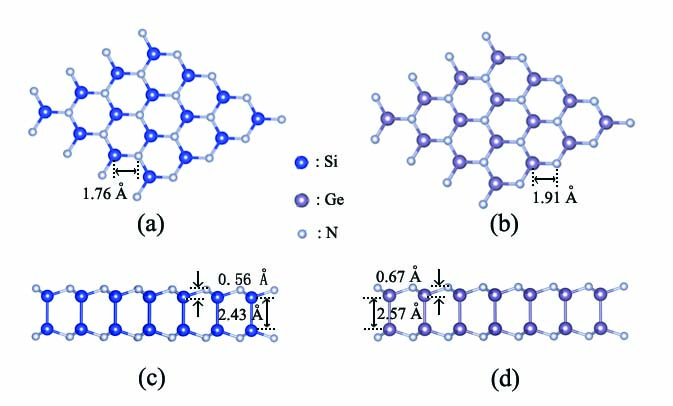}
\caption{(Color online). The optimized structures of ground-state 2D layered mononitrides SiN and GeN. (a) and (b) are for the top-views, and (c) and (d) are for the side-views.}
\label{fig:Figure1}
\end{figure}

Based on the comparison of 2D mononitrides and some relative compounds in bond length, it demonstrates that there are strong covalent bonds between the nearest-neighboring atoms in 2D layered SiN and GeN, and it can be confirmed by the isosurfaces of electron localization  function (ELF) plotted Fig. 2. The figure clearly shows that the electrons are distributed in the regions around the two nearest-neighboring atoms link, indicating the robust interaction between the two atoms. Besides, it obviously presents that all the bonds show $\sigma$ characteristic.

\begin{figure}[htbp]
\centering
\includegraphics[width=8.5cm]{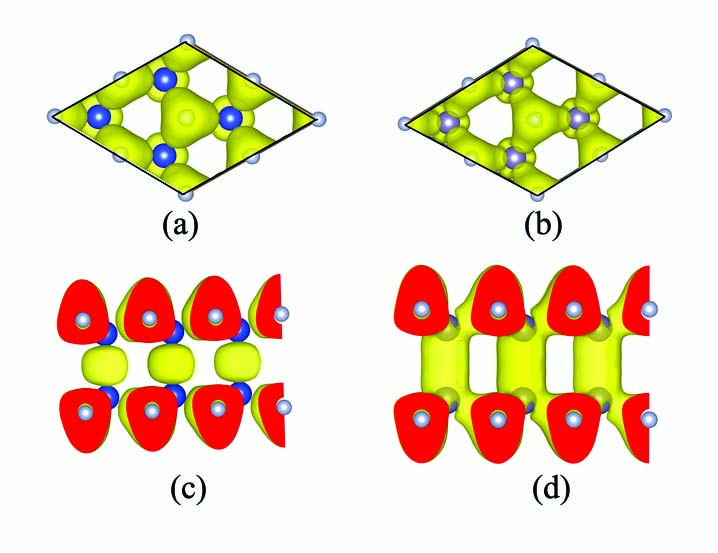}
\caption{(Color online). The isosurfaces of electron localization function (ELF) for 2D layered mononitrides SiN and GeN. (a) and (b) are for the top-views, and (c) and (d) are for the side-views.}
\label{fig:Figure2}
\end{figure}

To investigate the possibility of experimental synthesis of 2D layered mononitrides SiN and GeN, the thermodynamic stability of the ground-state structures is discussed through the cohesive energy. The cohesive energy is expressed as the follows:

\begin{equation}
\Delta E_{SiN}=(E_{Si}+E_{N}-E_{SiN})/2
\label{eq:fn}
\end{equation}

\begin{equation}
\Delta E_{GeN}=(E_{Ge}+E_{N}-E_{GeN})/2
\label{eq:fn}
\end{equation}

where E$_{SiN}$ and E$_{GeN}$ are the total energies of one chemical formula SiN and GeN, E$_{Si}$, E$_{Ge}$, and E$_{N}$ are the total energies of the isolated single Si, Ge, and N atoms, respectively. The calculated result reveals that $\Delta$E$_{SiN}$ and $\Delta$E$_{GeN}$ are 5.59 and 4.27 eV, respectively. These values are larger than 3.98 and 3.26 eV of silicene and gelicene calculated in this work. The relatively large cohesive energies of 2D SiN and GeN layers indicates their phases could exist stably.

Next, the dynamical properties of 2D SiN and GeN layers are studied by calculating the phonon dispersion, and the phonon dispersion curves are plotted in Fig. 3. If the curve shows no imaginary frequency, it indicates the structure is dynamically stable. Otherwise, the structure is dynamically unstable. The figure clearly presents that there is no imaginary frequency for the both mononitrides, telling that the two structures are both dynamically stable. Additionally, the highest frequencies of 2D SiN and GeN are around 800 and 700 cm$^{-1}$, respectively, much higher than those of silicene ($\sim$ 600 cm$^{-1}$) and germanene ($\sim$ 350 cm$^{-1}$) reported previously\cite{Cahangirov}. Such high frequencies suggest that strong bonds are formed between the two nearest neighboring atoms in SiN and GeN. Besides, the highest frequency of SiN is higher than that of GeN, corresponding to the fact that the bond lengthes of SiN are shorter than the counterparts of GeN.

\begin{figure}[htbp]
\centering
\includegraphics[width=8.5cm]{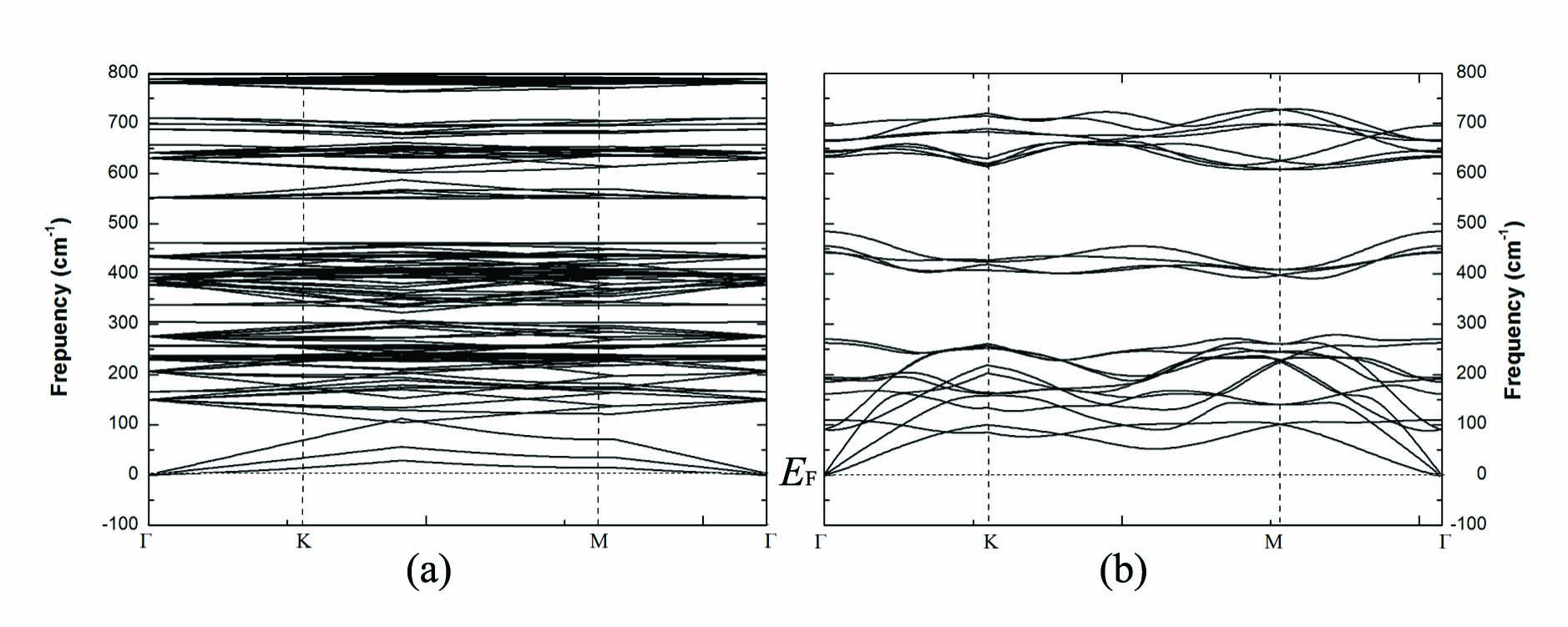}
\caption{(Color online). The phonon dispersion curves for 2D layered mononitrides SiN (a) and GeN (b).}
\label{fig:Figure3}
\end{figure}

To further evaluate the thermal stability of SiN and GeN, the ab initio molecular dynamics (AIMD) simulations are performed. During the calculations, a large 4$\times$4$\times$1 supercell based on primitive cell is employed, AIMD simulations are calculated using the NVT ensembles, the temperature controlled by the Nos\'{e}-Hoover method is ranged from 300 K to 2000 K, and the simulations last for 10 ps with a time step of 2.0 fs. Simulation snapshots of the last step for the two mononitrides at different temperature are described in Fig. 4. For SiN, it clearly shows that there is no breaking of the bonds and the original configuration is well kept even at high temperature of 2000 K, while the warp in the atomic plane enhances with increasing the temperature from the side view. For GeN, the performance is similar to that of SiN, except that the warp is much strong than that in SiN at the same temperature. When the temperature reaches 2000 K, some Ge atoms approach to breaking away from the atomic plane, but the original configuration is still remained mainly. These facts tell that the thermal stability of SiN is much stronger than that of GeN, originated from the higher cohesive energy of SiN than that of GeN. As a result, the above performance means that both layered SiN and GeN possess high thermal stability, indicating their potential applications even at extremely high temperature.

\begin{figure}[htbp]
\centering
\includegraphics[width=8.5cm]{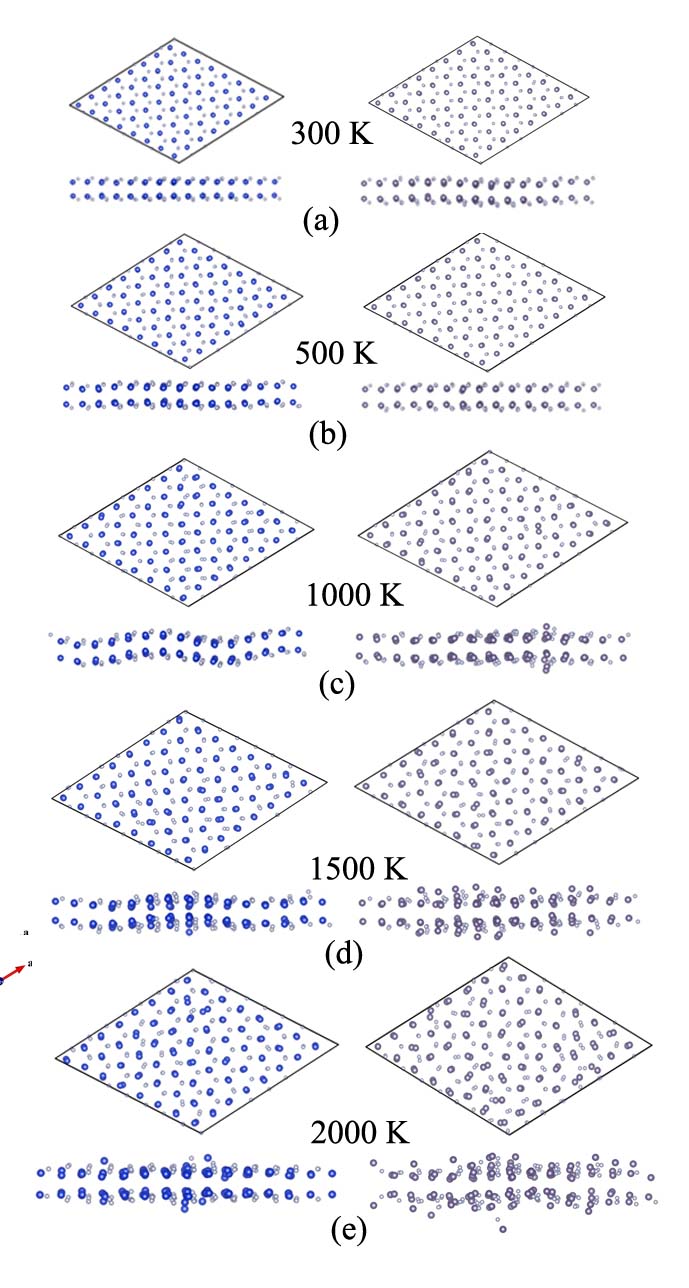}
\caption{(Color online). Snapshots of 2D layered mononitrides SiN (left column) and GeN (right column) equilibrium structures at (a) 300 K, (b) 500 K, (c) 1000 K, (d) 1500 K, and (e) 2000 K at the last step of 10 ps AIMD simulations.}
\label{fig:Figure4}
\end{figure}

Based on the stability of SiN and GeN, further investigation of their electronic properties is encouraged. The density of electronic states (DOS), and partial density of electronic states (PDOS) are calculated and described in Fig. 5. The total DOS in Figs. 5(a) and 5(c) clearly presents the semiconducting nature of the two layered mononitrides, and the energy gaps are 1.75 and 1.20 eV for SiN and GeN, respectively. Since the binding energy of valence electrons in Si is much stronger than that in Ge, the energy gap of SiN is $\sim$0.55 eV larger than that of GeN. In detail, the PDOS displays that there are many peaks of the Si/Ge and N states located at the same energy levels, revealing the strong interaction between two type atoms. This can be confirmed by ELF in Fig. 2 which exhibiting strong covalence bonds. Furthermore, the band structures are explored as well, as plotted in Figs. 5(b) and 5(d). It presents that SiN and GeN both behave as indirect band gap semiconductors, since the the conduction band minimum (CBM) and the valence band maximum (VBM) are situated at $\Gamma$ and \textit{K} reciprocal points, respectively. The projected density of electronic states of each atoms, as pictured from Fig. 5(e) to Fig. 5(h), detailedly unveils that VBM is mainly composed of Si/Ge \textit{p}$_{z}$ and N \textit{p}$_{z}$ orbitals, and CBM is mainly composed of Si/Ge \textit{s} and N \textit{p}$_{z}$ orbitals. Figure 6 plots the partial (band decomposed) charge density of valence and conduction band edges, and it shows that both CBM and VBM of the two layered compounds SiN and GeN are originated from the obitals of both two type atoms. The charge at VBM is localized around N atom and along Si-Si (Ge-Ge) bond, forming a $\sigma$ bond, and the charge at CBM is localized along Si-Si (Ge-Ge) with $\pi$ bond characteristic and some is around N atom.

\begin{figure}[htbp]
\centering
\includegraphics[width=8.5cm]{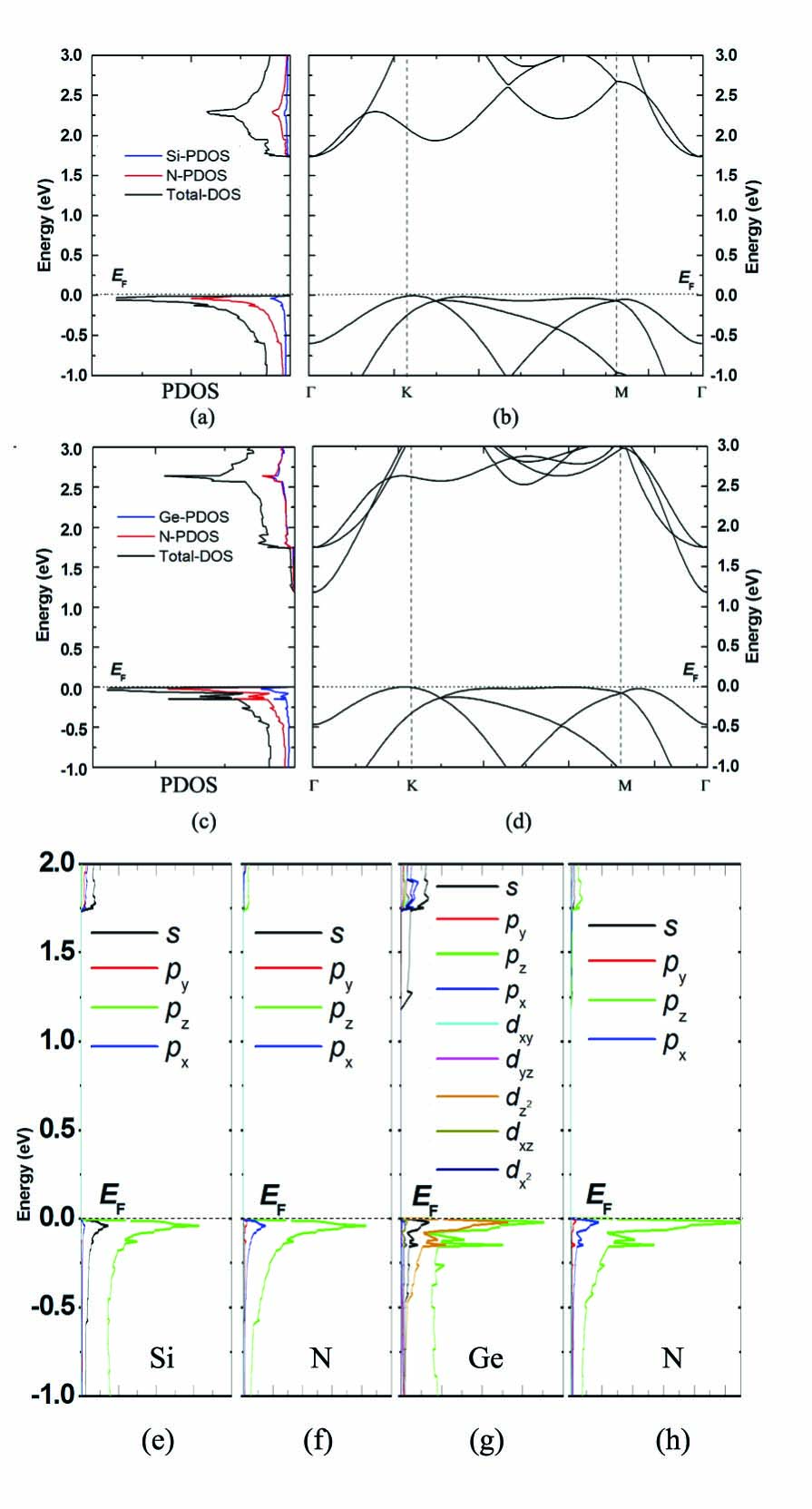}
\caption{(Color online). The density of electronic states (DOS) and partial density of electronic states (PDOS) for 2D layered mononitrides SiN (a) and GeN (c). The band structures for 2D layered mononitrides SiN (b) and GeN (d). The projected density of electronic states of Si (e) and N (f) in SiN, and Ge (g) and N (h) in GeN.}
\label{fig:Figure5}
\end{figure}

\begin{figure}[htbp]
\centering
\includegraphics[width=8.5cm]{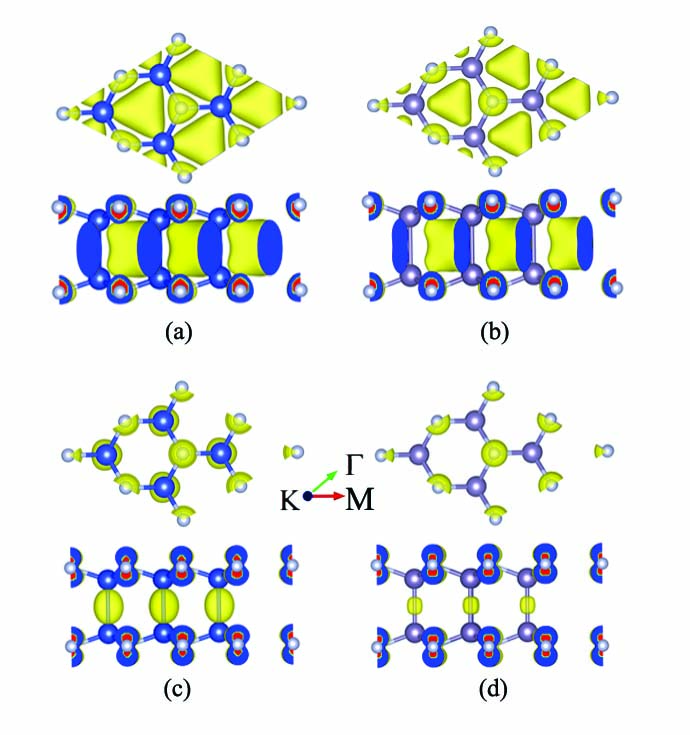}
\caption{(Color online). (a) and (b) are the partial (band decomposed) charge density of  conduction band edges for 2D layered mononitrides SiN and GeN, respectively. (c) and (d) are the partial (band decomposed) charge density of valence band edges for 2D layered mononitrides SiN and GeN, respectively.}
\label{fig:Figure6}
\end{figure}

Notably, the charge carrier transport property is so important for practical applications and thus is explored quantitatively. The phonon limited carrier mobility in 2D materials can be calculated by the definition\cite{Bruzzone,Takagi,Fiori,Qiao,Jia}

\begin{equation}
\mu_{2D}=\frac{e\hbar^{3}C_{2D}}{k_{B}Tm^{*}m_{a}(E_{l}^{i})^{2}}
\label{eq:fn}
\end{equation}

where e is the electron charge and $\hbar$ is Planck's constant divided by 2$\pi$, \textit{k}$_{B}$ is Boltzmann's constant, and \textit{T} is the temperature with 300 \textit{K} employed here. m$^{*}$ is the effective mass at CBM and VBM in the transport direction and defined as m$^{*}$=$\hbar$$^{2}$($\partial$$^{2}$E/$\partial$k$^{2}$)$^{-1}$, and m$_{a}$ is the average effective mass determined by m$_{a}$ = (m$^{*}$$_{K-M}$m$^{*}$$_{K-\Gamma}$)$^{1/2}$, \textit{C}$_{2D}$ is the elastic modulus of the longitudinal strain in the propagation directions of the longitudinal acoustic wave and expressed by (\textit{E}-\textit{E}$_{0}$)/\textit{S}$_{0}$=\textit{C}$_{2D}$($\Delta$\textit{l}/\textit{l}$_{0}$)$^{2}$/2, here \textit{E} is the total energy and \textit{S}$_{0}$ is the lattice volume at equilibrium for a 2D system. \textit{E}$_{l}$$^{i}$ with the expression \textit{E}$_{l}$$^{i}$=$\Delta$\textit{V}$_{i}$/($\Delta$\textit{l}/\textit{l}$_{0}$) represents the deformation potential constant of VBM for hole or CBM for electron along the transport direction. Here $\Delta$\textit{V}$_{i}$ is the energy change of the \textit{i}$^{th}$ band under proper cell compression and dilatation, \textit{l}$_{0}$ and $\Delta$\textit{l} are the lattice constant in the transport direction and the deformation of \textit{l}$_{0}$, respectively. According to this equation, the evaluated electron mobility is $\sim$0.888$\times$10$^{3}$ and $\sim$0.413$\times$10$^{3}$ cm$^{2}$V$^{-1}$s$^{-1}$ for SiN and GeN, respectively. The hole mobility in SiN is calculated as $\sim$0.191$\times$10$^{3}$ and $\sim$0.060$\times$10$^{3}$ cm$^{2}$V$^{-1}$s$^{-1}$ along K-$\Gamma$ and K-M directions, respectively, and it is $\sim$0.296$\times$10$^{2}$ and $\sim$0.069$\times$10$^{2}$ cm$^{2}$V$^{-1}$s$^{-1}$ for GeN along the same two directions, respectively. This carrier mobility anisotropy can be explained by partial charge density of valence and conduction band edges in Fig. 6. For VBM, the figure indicates that the charge spreads along K-$\Gamma$ direction rather than K-M direction, illustrating that the carrier transporting along K-$\Gamma$ direction is relatively easily than along K-M direction, since the carrier must cross the hexagonal hole along K-M direction. While for CBM, the charge fills the whole hexagonal hole, leading to the same electron mobility along different directions. Notably, these mobilities of the two 2D mononitrides are much higher than those of Boron Nitride nanoribbons ($\sim$58.80 cm$^{2}$V$^{-1}$s$^{-1}$)\cite{HZ} and MoS$_{2}$ ($\sim$3.0 cm$^{2}$V$^{-1}$s$^{-1}$)\cite{KS}, and have the same order of magnitude of $\sim$10$^{3}$ cm$^{2}$V$^{-1}$s$^{-1}$ in atomically thin InSe\cite{Denis} at room temperature. Such high carrier mobility demonstrates that 2D mononitrides SiN and GeN could be potential materials for high efficiency solar cell and so on.

\section{Conclusions}
In conclusion, via first principles calculations in combination with a swarm structure search method, 2D atomically thin mononitrides SiN and GeN are designed. Both layered mononitrides are constructed by two hexagonal single-atom-thick structures which linked by strong Si-Si or Ge-Ge bonds. The atomic configuration can be kept even at extremely high temperature up to 2000 K. The electronic calculation reveals that 2D layered SiN and GeN both exhibit semiconducting nature, and the band gaps are 1.75 and 1.20 eV, respectively. Excitedly, their carrier mobilities have a magnitude of $\sim$10$^{3}$ cm$^{2}$V$^{-1}$s$^{-1}$. This work opens a promising way for developing 2D atomically thin semiconducting materials which can match well with the previous substrates.

\vspace{1ex}
\begin{acknowledgments}
This work was supported by the National Natural Science Foundation of China (Grant nos. 11404168, 11304155, and 11374160). Y. Qian, R. Zhu, and Z. Du contributed equally to
this work.
\end{acknowledgments}

\end{document}